\documentclass[letterpaper]{article}
\usepackage{spconf,amsmath,amssymb,amsfonts,graphicx,cite,bm,booktabs}
\usepackage{microtype}
\usepackage{tikz}
\usetikzlibrary{arrows.meta, positioning}
\newcommand{\HH}{\mathsf H}
\newcommand{\CN}{\mathcal{CN}}
\newcommand{\tr}{\operatorname{tr}}
\newcommand{\clip}{\operatorname{clip}}

\title{Generative Large-Array Emulation for DOA Estimation in MIMO Radar via Conditional Diffusion}
\name{Adam Umra$^{\star}$, Aya Mostafa Ahmed$^{\dagger}$, Aydin Sezgin$^{\star}$%
\thanks{This work was supported by the German Research Foundation (DFG) under Project–ID 287022738 TRR 196 for Project S03.}}
\address{$^{\star}$Ruhr University Bochum, Germany\\
         $^{\dagger}$Robert Bosch GmbH, Germany}

\begin{document}
\maketitle
\begin{abstract}
Accurate direction-of-arrival (DOA) estimation from a small multiple-input multiple-output (MIMO) radar array is limited by its aperture, while adding antennas increases hardware cost. Snapshot-based array emulation reconstructs large-array observations before applying multiple signal classification (MUSIC), but reconstruction errors can distort the spectral peaks used for DOA estimation. We instead generate an idealized large-array MUSIC spectrum directly from small-array measurements. Coarray compression combines redundant virtual channels, and the resulting covariance and MUSIC spectrum condition a diffusion model. Consensus over peaks in multiple generated spectra yields the DOA estimates. Simulations with fluctuating targets show that the method outperforms a scene-matched snapshot-reconstruction network and an otherwise matched deterministic spectrum predictor. Small- and large-array MUSIC and their Cramér–Rao bounds provide reference comparisons. The method improves angle estimation using only the small array at inference.
\end{abstract}

\section{Introduction}
\label{sec:introduction}
Estimating target angles accurately with a limited number of antennas remains challenging. In colocated multiple-input multiple-output (MIMO) radar, the transmit and receive antennas form a virtual array whose aperture determines the angular resolution available for direction-of-arrival (DOA) estimation~\cite{li2007mimo}. Adding antennas can improve resolution, but increases hardware cost and complexity. Array emulation instead uses measurements from a smaller physical array to infer the angular information associated with a larger aperture, offering a signal-processing approach alongside hardware-based solutions such as reconfigurable surfaces~\cite{umra2026hardware}.

Existing array-emulation methods formulate this task as small-to-large-array snapshot reconstruction. Coupled dictionary learning was introduced in~\cite{miriya2020extrapolation}, followed by a deep neural network (DNN) mapping in~\cite{ahmed2021emulation}. Both subsequently apply multiple signal classification (MUSIC)~\cite{schmidt1986music} to the reconstructed observations. This formulation, however, optimizes an intermediate signal representation rather than the final quantity used for direction-of-arrival (DOA) estimation. The reconstructed snapshots must first be converted into a sample covariance matrix and then eigendecomposed to obtain the MUSIC spectrum. Consequently, a small element-wise reconstruction error may still perturb the estimated signal subspace or reverse the ordering of competing spectral peaks. Furthermore, the particular large-array noise realization is not determined by the observed small-array measurement. Reconstructing this nuisance component therefore consumes model capacity without contributing DOA information. These limitations motivate predicting the large-array MUSIC spectrum directly.

Direct spectrum estimation has been investigated independently of array emulation. The Super-Resolution Angular Spectra Estimation Network (SR-SPECNet) exploits one-dimensional radar structure~\cite{zheng2025srspecnet}, while the Spatial Spectrum Network, $(\mathrm{SP})^2$-Net, estimates a high-resolution spectrum from one snapshot~\cite{berman2025sp2}. In array emulation, finite small-array measurements leave uncertainty about the unobserved large-array spectrum. This motivates a generative model that can produce candidate spectra for a given observation. Diffusion probabilistic models provide such a mechanism by learning to reverse a gradual noising process~\cite{ho2020ddpm}. They have been applied to signal denoising for DOA estimation~\cite{qian2026doa} and to covariance recovery~\cite{liu2026covariance}. Neither application generates a large-array MUSIC spectrum conditioned on small-array covariance and MUSIC information.

We therefore propose \emph{conditional spectrum diffusion}, a direct small-to-large-array spectrum mapping. Starting from the small-array measurements, we coherently combine redundant virtual channels and compute a trace-normalized covariance matrix together with its MUSIC spectrum. A lossless encoding of the covariance and the normalized spectrum jointly condition a diffusion model trained to generate the ideal MUSIC spectrum of the corresponding large array. The covariance preserves complex spatial and multi-target relationships, while the MUSIC spectrum provides an explicit angular-domain representation. During inference, only the small-array measurements are required: the model transforms Gaussian noise into several candidate large-array spectra through a short sequence of denoising steps, after which peak consensus combines the candidates to obtain the final DOA estimates. 

The contributions are thus threefold: 1) direct small-to-large-array spectrum emulation from covariance and MUSIC information, 2) a compact conditional diffusion method with peak-weighted training and five-step sampling, and 3) a scene-matched evaluation against the state of the art~\cite{ahmed2021emulation} and a deterministic ablation on independent continuous-angle test data.

\section{System Model}
\label{sec:model}
\subsection{Matched-filter observation model}
Consider a colocated MIMO radar with $M_i$ transmit (TX) and $N_i$ receive (RX) antennas, where $i\in\{s,L\}$ denotes the small or large setup. Both TX and RX antennas form half-wavelength-spaced uniform linear arrays. Each TX antenna emits an orthogonal narrowband waveform comprising $P$ nonoverlapping pulses. We assume $K$ far-field point targets, with DOAs $\boldsymbol\theta=[\theta_1,\ldots,\theta_K]^T$. After matched filtering, the observation for pulse $p$ is
\begin{equation}
 \mathbf y_{i,p}=\sum_{k=1}^{K}\alpha_{k,p}\mathbf a_i(\theta_k)
 +\mathbf n_{i,p}
 =\mathbf A_i(\boldsymbol\theta)\mathbf x_p+\mathbf n_{i,p}.
 \label{eq:received}
\end{equation}
Here $\alpha_{k,p}$ is the radar cross section of target $k$ during pulse $p$, $\mathbf x_p=[\alpha_{1,p},\ldots,\alpha_{K,p}]^T$, and $\mathbf A_i(\boldsymbol\theta)=[\mathbf a_i(\theta_1),\ldots,\mathbf a_i(\theta_K)]$. The virtual-array steering vector is $\mathbf a_i(\theta)=\mathbf a_{r,i}(\theta)\otimes\mathbf a_{t,i}(\theta)$ with $\mathbf a_{t,i}(\theta)=[e^{j\pi m\sin\theta}]_{m=0}^{{M_i}-1}$ and $\mathbf a_{r,i}(\theta)=[e^{j\pi n\sin\theta}]_{n=0}^{{N_i}-1}$ being the transmit and receive steering vectors, respectively. We use a Swerling-II model, $\alpha_{k,p}\sim\CN(0,1)$ independently across targets and pulses~\cite{swerling1960}, and noise $\mathbf n_{i,p}\sim\CN(\mathbf0,\sigma_n^2\mathbf I)$. Stacking the matched-filter outputs gives
\begin{equation}
 \mathbf Y_i=[\mathbf y_{i,1},\ldots,\mathbf y_{i,P}]
 =\mathbf A_i(\boldsymbol\theta)\mathbf X+\mathbf N_i
 \in\mathbb C^{M_iN_i\times P}.
 \label{eq:matrix_observation}
\end{equation}
During paired-data generation, the two setups observe the same target realization $\mathbf X$, while $\mathbf N_s$ and $\mathbf N_L$ are drawn independently, so that the small-array observation contains no information about the particular large-array noise realization. Before forming the covariance matrices used for MUSIC, we combine in the following the redundant virtual channels in the matched-filter outputs.

\subsection{Coarray compression and MUSIC}
$M_iN_i$ TX--RX pairs form a redundant sum coarray with virtual positions $v=m+n$ but only $V_i=M_i+N_i-1$ distinct elements~\cite{chen2008minimum,ma2022nystrom}. Let $\mathbf S_i\in\{0,1\}^{M_iN_i\times V_i}$ assign each channel to its virtual position and let $\mathbf D_i=\mathbf S_i^T\mathbf S_i$ contain the multiplicities. We coherently combine repeated positions as
\begin{equation}
 \mathbf Z_i=\mathbf C_i\mathbf Y_i\quad \text{with} \quad \mathbf C_i=\mathbf D_i^{-1/2}\mathbf S_i^T.
 \label{eq:compression}
\end{equation}
This is called coarray compression: equal-$v$ channels have identical steering phase, so it discards only noise orthogonal to the modeled signal space. This reduces the number of covariance-matrix entries from $(M_iN_i)^2$ to $V_i^2$. The normalization by $\mathbf D_i^{-1/2}$ ensures that $\mathbf C_i\mathbf C_i^T=\mathbf I_{V_i}$, preserving white noise under the transformation, as required for MUSIC. Using these compressed snapshots, we compute
\begin{equation}
 \mathbf R_i=\frac{1}{P}\mathbf Z_i\mathbf Z_i^{\HH}.
 \label{eq:covariance}
\end{equation}
To remove the irrelevant overall power scale and provide consistently scaled network inputs, we normalize the covariance to unit trace $\overline{\mathbf R}_i=\mathbf R_i \slash \tr(\mathbf R_i)$.
This scaling preserves its eigenvectors and eigenvalue ratios and therefore leaves the MUSIC subspaces unchanged. The resulting steering vector of the compressed array is  $\mathbf b_i=\mathbf C_i\mathbf a_i$ and has entries
$[\mathbf b_i(\theta)]_v=\sqrt{[\mathbf D_i]_{vv}}e^{j\pi v\sin\theta}$, $v=0,\ldots,V_i-1$.
Assuming $K$ is known, let $\mathbf U_n$ contain the $V_i-K$ eigenvectors associated with the smallest eigenvalues of $\overline{\mathbf R}_i$~\cite{schmidt1986music}. Thus, the MUSIC pseudospectrum is
\begin{equation}
 P_i(\theta)=
 \frac{\mathbf b_i^{\HH}(\theta)\mathbf b_i(\theta)}
 {\mathbf b_i^{\HH}(\theta)\mathbf U_n\mathbf U_n^{\HH}\mathbf b_i(\theta)}.
 \label{eq:music}
\end{equation}

\section{Conditional Spectrum Diffusion}
\label{sec:method}
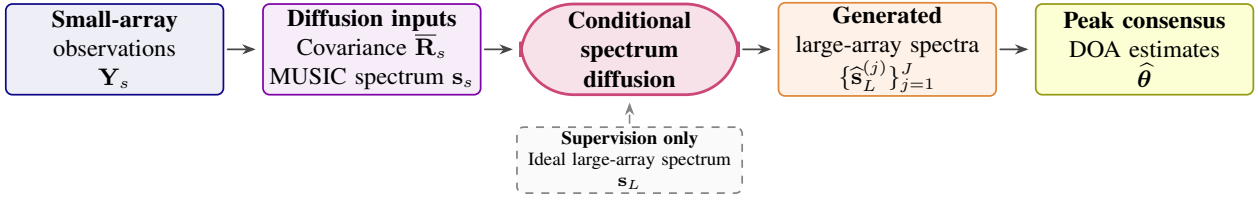
\begin{figure*}[t]
 \centering
 \definecolor{mycolor1}{HTML}{0d0887}
\definecolor{mycolor2}{HTML}{7e03a8}
\definecolor{mycolor3}{HTML}{cc4778}
\definecolor{mycolor4}{HTML}{f89540}
\definecolor{mycolor5}{HTML}{f0f921}

\begin{tikzpicture}[
    font=\small,
    node distance=5mm,
    stage/.style={
        draw,
        line width=0.65pt,
        rounded corners=2.5pt,
        align=center,
        text width=27mm,
        minimum height=12mm,
        inner xsep=2.5pt,
        inner ysep=1.5pt
    },
    diffusion stage/.style={
        stage,
        draw=mycolor3,
        fill=mycolor3!16,
        line width=1.05pt,
        rounded corners=7mm
    },
    flow/.style={
        -{Stealth[length=1.8mm,width=1.35mm]},
        draw=black!72,
        line width=0.65pt,
        shorten <=1.2pt,
        shorten >=2.4pt,
        line cap=round
    },
    supervision/.style={
        -{Stealth[length=1.7mm,width=1.25mm]},
        draw=black!48,
        line width=0.6pt,
        dashed,
        shorten <=1.2pt,
        shorten >=2.4pt,
        line cap=round
    }
]
    \node[stage,draw=mycolor1,fill=mycolor1!7] (measurements)
        {\textbf{Small-array}\\observations\\$\mathbf Y_s$};

    \node[stage,draw=mycolor2,fill=mycolor2!7,
        right=of measurements] (features)
        {\textbf{Diffusion inputs}\\Covariance $\overline{\mathbf R}_s$\\MUSIC spectrum $\mathbf s_s$};

    \node[diffusion stage,
        right=of features] (diffusion)
        {\textbf{Conditional}\\\textbf{spectrum}\\\textbf{diffusion}};

    \node[stage,draw=mycolor4!90!black,fill=mycolor4!12,
        right=of diffusion] (candidates)
        {\textbf{Generated}\\large-array spectra\\$\{\widehat{\mathbf s}_L^{(j)}\}_{j=1}^{J}$};

    \node[stage,draw=mycolor5!62!black,fill=mycolor5!20,
        right=of candidates] (output)
        {\textbf{Peak consensus}\\DOA estimates\\$\widehat{\boldsymbol\theta}$};

    \node[stage,dashed,draw=black!48,fill=black!2,text width=28mm,
        minimum height=8mm,font=\scriptsize,below=4mm of diffusion] (target)
        {\textbf{Supervision only}\\Ideal large-array spectrum\\$\mathbf s_L$};

    \draw[flow] (measurements.east) -- (features.west);
    \draw[flow] (features.east) -- (diffusion.west);
    \draw[flow] (diffusion.east) -- (candidates.west);
    \draw[flow] (candidates.east) -- (output.west);
    \draw[supervision] (target.north) -- (diffusion.south);
\end{tikzpicture}
 \caption{Conditional spectrum diffusion. The small-array covariance and MUSIC spectrum condition the model, while the ideal spectrum is required only during training.}
 \label{fig:architecture}
\end{figure*}

The previous section provides the compressed small-array covariance $\overline{\mathbf R}_s$ and its MUSIC spectrum $P_s$. Because direct peak detection on $P_s$ is limited by the small aperture, we use both quantities to estimate the spectrum of a larger array and then extract its peaks. Unlike the snapshot reconstruction in~\cite{ahmed2021emulation}, the proposed method predicts the large-array spectrum instead of reconstructing every large-array snapshot. The covariance retains the complex spatial relations, while $P_s$ explicitly locates the angular energy. Fig.~\ref{fig:architecture} summarizes the method.

\subsection{Network input and training target}
Following common real-valued representations of complex covariance inputs~\cite{elbir2020deepmusic}, we encode the Hermitian matrix \(\overline{\mathbf R}_s\) without redundancy as \(\mathbf r_s\in\mathbb R^{V_s^2}\), comprising its real diagonal and the real and imaginary parts of its strict upper triangle. The network receives $\mathbf{r}_s$ and $\mathbf s_s$, where $\mathbf s_s$ is the sampled normalized small-array spectrum of $P_s$. Thus, the condition $\mathbf c=(\mathbf r_s,\mathbf s_s)$ is obtained entirely from $\mathbf Y_s$.

During training, the corresponding target scene is also evaluated for the large array. Its ensemble covariance is
\begin{equation}
 \mathbf R_{L}
 =\mathbf B_L\mathbf B_L^{\HH}+\sigma_n^2\mathbf I_{V_L},
 \quad
 \mathbf B_L=[\mathbf b_L(\theta_1),\ldots,\mathbf b_L(\theta_K)].
 \label{eq:ideal}
\end{equation}
The target $\mathbf s_{L}$ is the normalized MUSIC spectrum of this covariance. The large-array spectrum is \emph{ideal} because it excludes finite-snapshot fluctuations. It is available only for supervision.

\subsection{Diffusion model and training}
The diffusion model learns to recover the clean large-array spectrum from a noisy version of it, conditioned on the small-array covariance encoding and normalized MUSIC spectrum. We first standardize the target as $\mathbf x_0=(\mathbf s_{L}-\boldsymbol\mu)\oslash\mathbf d$, using the training-set mean $\boldsymbol\mu$ and standard deviation $\mathbf d$. At a randomly selected diffusion level $t$, Gaussian noise is added according to
\begin{equation}
 \mathbf x_t=\sqrt{\bar\alpha_t}\mathbf x_0+
       \sqrt{1-\bar\alpha_t}\boldsymbol\epsilon,\qquad
 \boldsymbol\epsilon\sim\mathcal N(\mathbf0,\mathbf I_G).
 \label{eq:forward}
\end{equation}
A cosine schedule controls the noise level $\bar\alpha_t$~\cite{ho2020ddpm,nichol2021improved}. Given $\mathbf x_t$, $t$, and the small-array condition, the denoiser predicts the clean spectrum coordinates as
\begin{equation}
 \widehat{\mathbf x}_0=
 \sqrt{\bar\alpha_t}\mathbf x_t+
 f_\phi(\mathbf x_t,t,\mathbf r_s,\mathbf s_s).
 \label{eq:predict}
\end{equation}
The first term is a skip connection, while $f_\phi$ learns the required correction. A linear encoder maps $\mathbf r_s$ to a context vector, which is combined with a sinusoidal representation of $t$. A one-dimensional convolutional network processes $\mathbf x_t$ together with $\mathbf s_s$. Its residual blocks use feature-wise linear modulation (FiLM)~\cite{perez2018film} and dilated convolutions to relate peaks across different angular scales. The output is decoded as $\widehat{\mathbf s}_L=\clip(\boldsymbol\mu+\mathbf d\odot\widehat{\mathbf x}_0,0,1)$. To limit the spectral dynamic range, values more than $\Gamma$
decades below the maximum are clipped to a common floor before
rescaling.

We supervise the denoiser in both the standardized and decoded spectral domains by minimizing
\begin{equation}
 \mathcal L=\|\widehat{\mathbf x}_0-\mathbf x_0\|_2^2
 +\lambda_p\|\widehat{\mathbf s}_L-\mathbf s_L\|_{\mathbf W}^2,
 \label{eq:loss}
\end{equation}
where the diagonal weights in $\mathbf W$ equal $w_p>1$ close to a target and one elsewhere. The model therefore learns the full spectrum while emphasizing the DOA peaks. Target angles only define these training weights. We maintain an exponential moving average (EMA) of the parameters and select checkpoints by validation DOA mean squared error (MSE).

\subsection{Spectrum generation and DOA estimation}
At inference, the large-array target spectrum used during training is unavailable, so we generate candidate spectra conditioned on the observed small-array inputs $\mathbf{c}$. We start from Gaussian noise and use denoising diffusion implicit model (DDIM) sampling to traverse selected training noise levels from high noise to low noise. Skipping intermediate levels reduces the number of denoising steps at inference. At every level, the network estimates the clean spectrum, and DDIM sampling uses it to obtain the next, less noisy state. Repeating the procedure with several fixed noise initializations gives several candidate large-array spectra for the same observation.

For each candidate, we select and interpolate the $K$ strongest separated peaks. The $K$ largest peaks of the accumulated vote define $\widehat{\boldsymbol\theta}$. This consensus favors angle locations that recur across diffusion samples and avoids directly averaging spectra with different competing peaks.

\section{Results}
\label{sec:results}
\begin{figure*}[t]
 \centering
 \includegraphics[width=\textwidth]{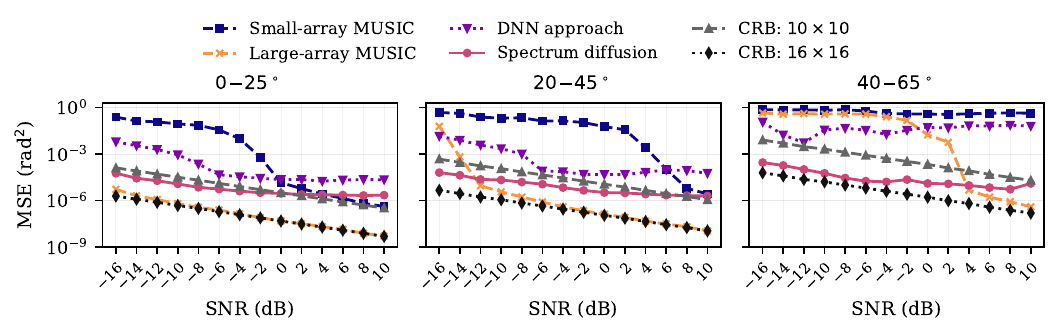}
 \caption{ DOA MSE for 200 physical scenes per noise level and range, with learned methods averaged. \emph{DNN approach} is the scene-matched linear-output snapshot DNN with full MUSIC. CRBs are scene-conditioned measurement-model references.}
 \label{fig:mse}
\end{figure*}

\subsection{Simulation and Evaluation Setup}
We use $(M_s,N_s)=(10,10)$ and $(M_L,N_L)=(16,16)$, yielding $V_s=19$ and $V_L=31$, with $K=4$ targets and $P=150$ pulses. The covariance and timestep encoders have dimensions 128 and 64. The width-32 spectral backbone has five FiLM blocks with dilations $1,2,4,8,16$, giving 144,577 parameters. We use 20 training levels, five DDIM levels, and four sampling trajectories. Furthermore, $w_p=10$ within $1^\circ$ of a target, $\lambda_p=1$, and the consensus bandwidth is $0.5^\circ$. The Adaptive Moment Estimation (Adam) optimizer with decoupled weight decay (AdamW)~\cite{loshchilov2019decoupledweightdecayregularization} uses a learning rate of $2\times10^{-4}$, batch size 64, and EMA decay 0.995 for at most 300 epochs.

Separate models cover $0^\circ$--$25^\circ$, $20^\circ$--$45^\circ$, and $40^\circ$--$65^\circ$. For each range and SNR, the 200 training scenes comprise equal numbers of $5^\circ$ configurations and four distinct random continuous angles with minimum spacing $1^\circ$. Validation uses 50 independently generated fixed-spacing scenes per SNR for model selection. After all models and settings were frozen, an independent test set was generated with 200 fixed-spacing scenes per SNR.

The main DNN baseline follows the snapshot-wise $200\rightarrow200\rightarrow200\rightarrow512\rightarrow512$ mapping in~\cite{ahmed2021emulation}, with feature-wise min--max normalization, rectified linear unit (ReLU) hidden layers, and Adam training for up to 150 epochs at batch size 120. Training uses the same scene observations as the proposed method: 30,000 raw snapshots per SNR.

All methods share the evaluation scenes and angular search grid. Covariance-based MUSIC selects peaks and refines them by continuous local maximization. neural spectra use quadratic interpolation, with consensus for diffusion. As a evaluation metric we use MSE
\begin{equation}
 \mathrm{MSE}=\frac{1}{QK}\sum_{j=1}^{Q}
   \min_{\pi\in\mathcal S_K}\sum_{k=1}^{K}
   (\widehat\theta_{j,\pi(k)}-\theta_{j,k})^2,
 \label{eq:mse}
\end{equation}
where $Q$ is the number of scenes, $\mathcal S_K$ contains all target permutations, and angles are in radians. The pooled root-mean-square error (RMSE) is $(180/\pi)\sqrt{\mathrm{MSE}}$, averaging squared errors before taking the root.

\subsection{DOA estimation performance}
Fig.~2 demonstrates the consistent advantage of spectrum diffusion across all three angular sectors. The proposed method achieves the lowest MSE among the learned approaches at every SNR, with an error that decreases steadily as the measurements improve. In contrast, the DNN approach reaches pronounced error floors, particularly in the challenging $40^\circ$--$65^\circ$ sector, where spectrum diffusion remains several orders of magnitude more accurate. The proposed method also outperforms small-array MUSIC over much of the evaluated SNR range and, in the two higher-angle sectors, remains below large-array MUSIC before its sharp high-SNR improvement. The Cramér–Rao bounds (CRBs) for the $10\times10$ and $16\times16$ arrays are plotted as reference lower bounds for unbiased measurement-based estimators. Overall, spectrum diffusion provides stable performance across SNRs and sectors while substantially narrowing the gap to physical large-array processing.

\subsection{Diffusion ablation}
Table~1 examines whether the diffusion procedure provides an advantage over direct conditional spectrum prediction. The deterministic ablation uses the same training data, conditioning inputs, target spectra, loss function, and spectral backbone as the proposed method, but omits diffusion noising, timestep conditioning, iterative sampling, and multi-sample consensus. It instead predicts the large-array spectrum in a single pass. The comparison therefore isolates the overall contribution of the diffusion-based generation procedure while keeping the underlying prediction task and supervision fixed. Whereas Fig.~2 reports MSE to enable comparison with the CRBs, Table~1 presents pooled RMSE in degrees for a more interpretable summary of angular estimation accuracy.

\begin{table}[h]
 \centering
 \caption{Diffusion ablation: RMSE, pooled across SNRs}
 \label{tab:ablation}
 \small
 \setlength{\tabcolsep}{2.5pt}
 \begin{tabular}{@{}lrrrrr@{}}
\toprule
Method & $0$--$25$ & $20$--$45$ & $40$--$65$ & All\\
\midrule
Direct predictor & 2.586 & 2.535 & 6.124 & 4.107\\
Spectrum diffusion & \textbf{0.188} & \textbf{0.219} & \textbf{0.425} & \textbf{0.297}\\
\bottomrule
\end{tabular}

\end{table}

Spectrum diffusion achieves a lower RMSE in every angular sector and reduces the pooled RMSE from $4.107^\circ$ to $0.297^\circ$, corresponding to a factor of $13.8$. The largest absolute improvement occurs in the $40^\circ$--$65^\circ$ sector, where the RMSE decreases from $6.124^\circ$ to $0.425^\circ$.

\section{Conclusion}
We presented conditional spectrum diffusion for estimating an ideal large-array spatial spectrum from small-array covariance and MUSIC information. Unlike snapshot-based array emulation, the proposed method directly learns the representation used for DOA estimation, avoiding the need to reconstruct unpredictable large-array noise. It substantially outperformed the baseline methods. These results demonstrate the effectiveness of diffusion-based spectrum generation and meaningfully narrow the performance gap between small- and large-array processing.

\clearpage
\bibliographystyle{IEEEbib}
\bibliography{bibliography}
\end{document}